\documentclass[twocolumn,showpacs,amsmath,amssymb,prl]{revtex4}
\usepackage{graphicx}
\usepackage{dcolumn}
\usepackage{bm}

\begin{document}
\input{epsf}

\title{Is a Classical Language Adequate in Assessing the Detectability \\
of the Redshifted 21cm Signal from the Early Universe?}

\author{Abraham Loeb}

\affiliation{Astronomy Department, Harvard University, 60 Garden St.,
Cambridge, MA 02138, USA}

\begin{abstract} 

The classical radiometer equation is commonly used to calculate the
detectability of the 21cm emission by diffuse cosmic hydrogen at high
redshifts. However, the classical description is only valid in the regime
where the occupation number of the photons in phase space is much larger
than unity and they collectively behave as a classical electromagnetic
field. At redshifts $z\lesssim 20$, the spin temperature of the
intergalactic gas is dictated by the radiation from galaxies and the
brightness temperature of the emitting gas is in the range of mK,
independently from the existence of the cosmic microwave background.  In
regions where the observed brightness temperature of the 21cm signal is
smaller than the observed photon energy, $h\nu=68(1+z)^{-1}$mK, the
occupation number of the signal photons is smaller than
unity. Nevertheless, the radiometer equation can still be used in this
regime because the weak signal is accompanied by a flood of foreground
photons with a high occupation number (involving the synchrotron Galactic
emission and the cosmic microwave background).  As the signal photons are
not individually distinguishable, the combined signal$+$foreground
population of photons has a high occupation number, thus justifying the use
of the radiometer equation.
\end{abstract}

\date{\today}
\maketitle

A black-body radiation field of temperature $T$ and a photon occupation
number $n=(\exp\{h\nu/kT\}-1)^{-1}$, is characterized by fluctuations in
the number of photons per quantum state with a root-mean-square amplitude
of~\cite{Moran},
\begin{equation}
\langle \Delta n_{\rm rms} \rangle=\sqrt{n(n+1)}=\sqrt{n^2+n} .
\label{flu}
\end{equation}  
At low occupation numbers ($n\ll 1$) or high photon energies, the
fluctuation amplitude is $(\Delta n_{\rm rms})\approx \sqrt{n}$, as
expected from Poisson statistics of independent photons.  This regime is
often encountered in optical, UV, or X-ray telescopes.  However, in the
regime where many photons occupy the same state ($n\gg 1$), the photons are
correlated and the $n^2$ term dominates over the Poisson term $n$ inside
the square root of Eq. (\ref{flu}).  Here, the ensemble of photons can be
described as a ``classical'' electromagnetic field with $\langle \Delta
n_{\rm rms}\rangle=n$.  The classical description applies to low photon
energies $h\nu\ll kT$ in the Rayleigh-Jeans (RJ) regime of the black-body
spectrum, and is naturally encountered in radio astronomy
\cite{Conf,Moran}.

The photon number fluctuations induce noise in a detector which is exposed
to the radiation field. For a detector that does not preserve the phase of
the incoming radiation, such as an ideal incoherent bolometer, the square
of the resulting {\it noise-equivalent-power} is given by
\cite{Moran,Mather,Griffin}
\begin{equation}
(\Delta P_{\rm rms})^2= 2\Omega A \int_0^\infty \left({2h\nu^3\over
c^2}\right)(\Delta n_{\rm rms})^2 h\nu d\nu ,
\label{nep}
\end{equation}
where $A$ is the collecting area of the detector and $\Omega$ is the solid
angle that it views.  For $n\gg 1$, Eqs. (\ref{flu}) and (\ref{nep}) can be
combined to get the noise level of a measurement over a narrow band of
frequencies $\Delta \nu$ centered on a frequency $\nu=c/\lambda$, $(\Delta
P_{\rm rms}) = 2 (A\Omega/\lambda^2)^{1/2}(\Delta \nu)^{1/2}(kT)$. This
result has no reference to Planck's constant $h$ as expected for the
classical regime, and provides a noise power that is simply proportional to
the temperature as expected for RJ scaling (where for a simple antenna, the
pre-factor $A\Omega/\lambda^2\sim 1$). However if $n\ll 1$, the Poisson
term dominates and yields a very different result for the noise power
$(\Delta P_{\rm rms})$, which is lower by a factor of
$(h\nu/kT)\exp\{-h\nu/2kT\}$ than the expression in the $n\gg1$ regime.  In
this case, the resulting noise power is not simply proportional to the
temperature.  The precision of the measurement is limited by Poisson
statistics, i.e. by the square-root of the total number of photons that are
collected by the telescope. Coherent detectors, such as interferometers,
preserve the phase information about the incoming radiation and have a
minimum noise temperature of $h\nu/k$ from Heisenberg's uncertainty
principle \cite{Moran,Woody}. This minimum noise already places them on the
borderline between the quantum and classical regimes. Our subsequent
discussion will not be restricted to any particular type of detector, but
rather focus on the unavoidable noise introduced by photon counting
statistics in the incoming radiation field.

Recently, there has been extensive research on the feasibility of mapping
the three-dimensional distribution of diffuse cosmic hydrogen less than a
billion years after the big bang, through its resonant spin-flip transition
at a wavelength of 21cm \cite{Furlanetto,BarLoeb}. Several experiments are
currently being constructed (such as
MWA\footnote{http://www.haystack.mit.edu/ast/arrays/mwa/},
LOFAR\footnote{http://www.lofar.org/}, PAPER
\footnote{http://astro.berkeley.edu/~dbacker/EoR/},
21CMA\footnote{http://web.phys.cmu.edu/~past/}) and others are being
designed (SKA\footnote{http://www.skatelescope.org/}) to detect the
theoretically predicted emission signal.

In gauging the detectability of the redshifted 21cm signal, the so-called
{\it radiometer equation} is commonly used to assess the sensitivity of an
interferometric array of low-frequency dipole antennae \cite{Moran}.  The
radiometer equation provides the root-mean-square brightness temperature
fluctuations $\langle\Delta T^2\rangle^{1/2}$ of a phased array of dipoles
at a frequency interval $\Delta \nu$ centered on a frequency
$\nu=c/\lambda$, in terms of the system temperature $T_{\rm sys}$ and the
integration time $t_{\rm int}$,
\begin{equation}
\langle\Delta T^2\rangle^{1/2}=\kappa {T_{\rm sys}\over {\sqrt{t_{\rm
int}\Delta\nu}}},
\label{radiom}
\end{equation}
where $\kappa\gtrsim 1$ is a parameter that describes the overall
efficiency of the array (and is of order the inverse of the aperture
filling fraction \cite{Furlanetto}).  For an optimized
(``filled-aperture'') array of dipoles which are separated from each other
by $\sim \lambda/2$, the factor $\kappa$ is of order unity.  The noise
power provided by the radiometer equation is proportional to the system
temperature as expected from RJ scaling, and has no reference to Planck's
constant. It is therefore valid only in the classical regime.  In our
discussion we will assume that the system temperature is dominated by the
sky brightness temperature, which includes both the signal and its
astronomical foregrounds.  For the sake of generality, we will ignore the
technology-dependent noise introduced by the detector (be it coherent or
incoherent \cite{Moran,Moran2}), and instead focus on the non-reducible
effect of the astronomical sky.
Our discussion applies to an ideal instrument in which the sensitivity is
dominated by the statistics of photons from the sky and not by other
(e.g. thermal or instrumental) sources of noise within the detector
\cite{Mather,Griffin}.

The radiometer equation expresses the simple fact of Gaussian statistics
that the noise temperature can be reduced by the square root of the number
of independent samples of this noise \cite{Moran}. For a frequency band
$\Delta \nu$, each random sample lasts a period of time $\sim
\Delta\nu^{-1}$. If many photons occupy the same volume of space and arrive
within the same sample time, then each random sampling of the noise would
contain many photons. This would imply that the noise properties are not
dictated by individual photons but rather by groups of photons arriving
within the same time interval $\sim \Delta\nu^{-1}$ and occupying the same
region in phase space. Such groups of overlapping photon wavepackets can be
described as a classical electromagnetic wave, whose fluctuation properties
are not associated with the statistics of uncorrelated photons. The photons
within each sample are correlated with each other because they occupy the
same volume in phase space, i.e. the same quantum state.

The radiometer equation is derived using the classical language of
electromagnetic fields \cite{Moran}. As explained above, this language is
adequate in the regime where many photons occupy each quantum state in
phase space. This regime, in which photons of a given frequency are
correlated with each other, is genuinely different from the regime often
encountered in optical, UV, or X-ray observations of astronomical sources,
for which Poisson fluctuations of uncorrelated photons determine the
signal-to-noise ratio \cite{Conf}.  When the occupation number is high,
there are multiple copies of each photon at a given frequency and arrival
time, and the detection of one photon by a particular dipole does not
preclude other dipoles from simultaneously detecting another photon at the
same quantum state. The noise properties are dictated by the number of
random samples, each of duration $\sim \Delta\nu^{-1}$, and not by the
number of detected photons.  This gives no dependence of the noise in
Eq. (\ref{radiom}) on the telescope diameter (for an optimized array with
$\kappa$ of order unity), in contrast with the dependence encountered in
optical or X-ray observations, where Poisson statistics of uncorrelated
photons with different arrival times implies that the fractional noise
amplitude is inversely proportional to the square root of the number of
detected photons or the square root of the collecting area of the
telescope.

We start our discussion by focusing on the signal photons because the
foregrounds can in principle be reduced. For example, Galactic synchrotron
radiation is expected to be highly polarized and therefore an appropriate
polarization filter can remove its polarized component. (A fully polarized
foreground could be fully removed while losing only 50\% of the unpolarized
\cite{Babich,CF} signal photons.) Any such filter should be applied to
small patches of the sky over which the foreground polarization is nearly
uniform.  The polarization filter needs to be rotated as a function of
frequency because of Faraday rotation by the interstellar magnetic field.
in the intervening ionized gas. Such a removal could be of great practical
value, since a reduction of the foreground brightness by a factor $f$ would
decrease by a factor of $f^2$ the integration time $t_{\rm int}$ required
to achieve a particular noise level $\langle\Delta T^2\rangle^{1/2}$ in
Eq. (\ref{radiom}). For now, let us ignore the foregrounds.

The number of photons occupying the same quantum state in phase space
around a frequency $\nu$ defines the photon occupation number (or the
dimensionless ``phase space density'' in $h^3$ bins), $n$. It is related to
the spectral intensity, $I_\nu$ (which has units of ${\rm
ergs~cm^{-2}~s^{-1}~Hz^{-1}~ster^{-1}}$), through the identity $n\equiv
(c^2/2h\nu^3)I_\nu$, where $h$ is Planck's constant.  The brightness
temperature of a radiation field is defined in radio astronomy through the
RJ relation, $kT_b=(\lambda^2/2)I_\nu$ (even when $I_\nu$ is not in the RJ
regime), yielding the simple result,
\begin{equation}
n={kT_b\over h\nu} .
\label{occ}
\end{equation}

At the redshifts of interest for the current generation of 21cm
experiments, $z\lesssim 20$, the excitation (spin) temperature of the
neutral intergalactic hydrogen is raised well above the temperature of the
cosmic microwave background (CMB) by the cumulative X-ray and Ly$\alpha$
radiation from galaxies \cite{Pritchard}.
First, we would like to demonstrate that in this regime, the desired
signal, i.e.  the 21cm brightness temperature of the gas, is
independent of the CMB and would have the same value even if the CMB
did not exist. This is an important point of principle, since the CMB
is characterized by $n\gg1$ but the signal that we are after
could be characterized by $n<1$. The radiative transfer equation
for a spectral line \cite{RL} reads,
\begin{equation}
{dI_\nu\over ds}={\phi(\nu) h\nu\over 4\pi}\left[n_2A_{21} -
\left(n_1B_{12} -n_2B_{21}\right)I_\nu\right],
\label{rad}
\end{equation}
where $ds$ is a line element, $\phi(\nu)$ is the line profile function
normalized by $\int \phi(\nu)d\nu=1$ (with an amplitude of order the
inverse of the frequency width of the line), subscripts 1 and 2 denote the
lower and upper levels, $n$ denotes the number density of atoms at the
different levels, and $A$ and $B$ are the Einstein coefficients for the
transition between these levels. We can then make use of the standard
relations \cite{RL}: $B_{21}=(g_1/g_2)B_{12}$ and
$B_{12}=(g_2/g_1)A_{21}n/I_\nu$, where $g$ is the spin degeneracy factor of
each state. For the 21cm transition, $A_{21}=2.85\times 10^{-15} {\rm
s^{-1}}$ and $g_2/g_1=3$ \cite{Field}.  The relative populations of
hydrogen atoms in the two spin states defines the so-called spin
temperature, $T_s$, through the relation, $(n_2/n_1)=
(g_2/g_1)\exp\{-E/kT_s\}$, where $E/k=68$ mK is the transition energy. In
the regime of interest, $E/k$ is much smaller than the CMB temperature
$T_{\rm cmb}$ as well as the spin temperature $T_s$, and so all related
exponentials can be expanded to leading order.  We may also replace $I_\nu$
with $2kT_b/\lambda^2$.  By substituting all the above relations in
Eq. (\ref{rad}), we get the observed brightness temperature $T_b$ in terms
of the optical depth in the 21cm line $\tau \ll1$,
\begin{equation}
T_b = {\tau \over (1+z)} \left(T_s- {En \over k}\right) .
\label{tbdt}
\end{equation}
In an expanding Universe with a uniform hydrogen number density $N_{\rm HI}$ and
with a velocity gradient equal to the Hubble parameter $H$,
\begin{equation}
\tau={3\over 32\pi}{h^3c^3 A_{21}\over E^2}{N_{\rm HI}\over H kT_s}.
\label{opt}
\end{equation}

In the presence of the CMB, $(En/k)\approx T_{\rm cmb}$, and the
right-hand-side of Eq. (\ref{tbdt}) is proportional to $(T_s-T_{\rm cmb})$,
yielding an emission signal if $T_s>T_{\rm cmb}$, as expected. But more
generally, Eqs. (\ref{tbdt}) and (\ref{opt}) imply that as long as $T_s\gg
T_{\rm cmb}$, the magnitude of the emission signal $T_b$ is not dependent
on the existence of the CMB (or even the particular value of $T_s$, as
$\tau\propto T_s^{-1}$).  Equations (\ref{tbdt}) and (\ref{opt}) were
already derived before \cite{Furlanetto,Tozzi,BL}.

Allowing for a small-amplitude density fluctuation and its
corresponding peculiar velocity modifies slightly the above expression
for the optical depth.  For a large-scale spherical region of a mean
over-density $\delta$ and a mean neutral fraction $X_{\rm HI}$ at a
redshift $z$, the observed mean brightness temperature is
\cite{Furlanetto,Tozzi,BL},
\begin{equation}
T_b=9{\rm mK}~ X_{\rm HI} \left(1+ {4\over 3}\delta\right)
\left(1+z\right)^{1/2} ,
\label{br}
\end{equation}
where we have adopted the standard values for the cosmological parameters
\cite{WMA}. The factor of ${4\over 3}$ is obtained from angular
averaging of $(1+\cos^2\theta)$, in which the $\theta$-dependent term
results from the line-of-sight gradient of the line-of-sight peculiar
velocity in the denominator of $\tau$, with $\theta$ being the angle
between the ${\bf k}$-vector of contributing Fourier modes and the
line-of-sight \cite{BL}.  Numerical simulations of reionization indicate
that the brightness temperature fluctuations in the 21cm emission of
hydrogen at $6\lesssim z\lesssim 20$ are $\lesssim 20$mK
\cite{McQuinn,Trac,Iliev}. At redshifts $z\lesssim 6$, the observed neutral
fraction by mass (mainly residing in damped Ly$\alpha$ absorbers) is
$X_{\rm HI}\approx 3\%$ and so on large scales $T_b\approx 0.3{\rm
mK}(1+{4\over 3}\delta) (1+z)^{1/2}$ \cite{WL}. In comparison, the observed
photon energy of the 21cm transition is $h\nu/k=68(1+z)^{-1}$mK.

From Eqs. (\ref{occ}) and (\ref{br}), we find that $n < 1$ for
cosmological regions in which
\begin{equation} 
X_{\rm HI}< 0.4\left(1+{4\over 3}\delta\right)^{-1}
\left({1+z \over 7}\right)^{-3/2},
\end{equation}
thus invalidating the use of the classical radiometer equation for the {\it
signal photons} on their own in this regime. For example, the detectable
21cm emission after reionization (smoothed over scales of tens of comoving
Mpc) \cite{WL} is entirely in this regime.  If the 21cm emission signal was
the only source of photons on the sky, then its detectability in the regime
where $n<1$ should have been calculated based on Poisson fluctuations
in photon counting statistics and {\it not} the radiometer equation.

However, the 21cm signal is accompanied by unavoidable foreground
photons with $n\gg1$.  The foregrounds obviously include the
CMB, for which
\begin{equation} 
n={1\over [{e^{h\nu/kT_{\rm cmb}}-1}]}\approx {kT_{\rm
cmb}\over h\nu} =40.1 \times (1+z), 
\end{equation} 
where $T_{\rm cmb}=2.73{\rm K}$ is the CMB temperature
\footnote{The sky brightness must be in the classical regime when cosmic
hydrogen appears in 21cm absorption at redshifts $z\gtrsim 30$ \cite{LZ},
because the CMB is required for this signal to exist.}.  In addition, the
Galactic synchrotron foreground has a brightness temperature of $T_{\rm
syn}\sim 140{\rm K} (\nu/200{\rm MHz})^{-2.6}$, yielding a flood of photons
with an occupation number $n\approx 1.4\times10^4 (\nu/200{\rm
MHz})^{-3.6}$. Although the polarized component of the synchrotron
foreground can in principle be removed by a polarization filter, the
unpolarized CMB component cannot be separated from the signal.  The
combination of the signal and residual foregrounds results in a population
of photons with a high occupation number that can be adequately described
as a classical electromagnetic field. Since the unpolarized signal photons
are not labeled by a quantum-mechanical tag that distinguishes them from
the unpolarized CMB photons, the detectability of even a weak signal is
adequately described by the classical language of the radiometer equation.

We therefore conclude that the use of the radiometer equation for gauging
the detectability of the 21cm signal is justified by the flood of noise
photons from the sky that accompanies the seeked-after signal. It is
important to keep this point in mind when considering the detectability of
other spectral lines \cite{Hogan,Loeb,Sethi, Sunyaev} or continuum emission
\cite{Loeb2,Cooray} from the diffuse cosmic gas, especially at wavelengths
shorter than a millimeter -- where the Wien tail of the CMB or other
foregrounds are not as bright.

The {\it Cosmic Background Explorer (COBE)} did already operate in the
regime where $h\nu/kT_{\rm cmb}>1$, but the high temperature of its
instruments introduced thermal noise 
\footnote{http://lambda.gsfc.nasa.gov/product/cobe/dmr\_exsup.cfm} that
brought it to the regime of $n\gg 1$ where the classical radiometer
equation applies. The performance of future CMB detectors that will be
cooled to temperatures $T< h\nu/k$ and will observe the CMB at frequencies
$\nu>kT_{\rm cmb}/h$ for which $n\ll 1$, will need to be based the
quantum-mechanical modification of the classical radiometer equation where
the $n$ term dominates over the $n^2$ term in Eq. (\ref{flu}).

\bigskip
\bigskip
\paragraph*{Acknowledgments.}
I thank Ron Ekers, Matt McQuinn, Jim Moran, George Rybicki, and Matias
Zaldarriaga, for helpful discussions. I am also grateful to the
Harvard-Australia foundation for an award that allowed me to write this
paper during an inspiring trip to Australia.

\end{document}